\documentclass[12pt,a4paper]{article}
\usepackage{amsfonts,latexsym}
\usepackage{graphicx,color}
\usepackage{dcolumn}
\usepackage{graphicx}
\usepackage{amsmath}
\usepackage{amsfonts}
\usepackage{amssymb}

\renewcommand{\thefootnote}{\fnsymbol{footnote}}

\begin{document}

\vspace{12mm}

\begin{center}
{{{\Large {\bf Quasinormal modes  of scalarized  black holes in the Einstein-Maxwell-Scalar theory}}}}\\[10mm]

{Yun Soo Myung$^a$\footnote{e-mail address: ysmyung@inje.ac.kr} and De-Cheng Zou$^{a,b}$\footnote{e-mail address: dczou@yzu.edu.cn}}\\[8mm]

{${}^a$Institute of Basic Sciences and Department  of Computer Simulation, Inje University Gimhae 50834, Korea\\[0pt] }

{${}^b$Center for Gravitation and Cosmology and College of Physical Science and Technology, Yangzhou University, Yangzhou 225009, China\\[0pt]}
\end{center}
\vspace{2mm}

\begin{abstract}
We perform the stability analysis on scalarized charged  black holes  in the Einstein-Maxwell-Scalar (EMS) theory  by computing quasinormal mode spectrum.
It is noted that the appearance of these  black holes with scalar hair is closely related to the instability of Reissner-Nordstr\"{o}m black holes without scalar hair in the EMS theory.
The scalarized charged black hole solutions are classified by the order number of $n=0,1,2,\cdots$, where $n=0$ is called the fundamental branch and $n=1,2,\cdots$ denote the $n$ excited branches.
Here, we show that  the $n=1,2$ excited black holes are unstable against  against the $s(l=0)$-mode scalar perturbation, while the $n=0$  black hole is stable against all scalar-vector-tensor perturbations. This is consistent with other scalarized black holes without charge found in the Einstein-Scalar-Gauss-Bonnet theory.

\end{abstract}
\vspace{5mm}

\vspace{1.5cm}

\hspace{11.5cm}
\newpage
\renewcommand{\thefootnote}{\arabic{footnote}}
\setcounter{footnote}{0}


\section{Introduction}
Recently, a scalarization of the Reissner-Nordstr\"{o}m (RN) black holes  was investigated  in the
Einstein-Maxwell-scalar (EMS) theory which is a simpler theory than the Einstein-Scalar-Gauss-Bonnet-scalar (ESGB) theory~\cite{Herdeiro:2018wub}.
Here, $q=Q/M$ may increase beyond unity,  compared to $0<q\le1$ for the RN black hole.
The EMS theory is a second-order theory which  includes three  propagating modes of scalar, vector, and tensor.
In this case, the instability of RN black hole was determined solely by the linearized scalar
equation because the RN black hole is stable against tensor-vector perturbations theory~\cite{Zerilli:1974ai,Moncrief:1974gw,Moncrief:1974ng,Moncrief:1975sb}.
It is shown that the appearance of the scalarized charged  black hole is closely associated with the Gregory-Laflamme (GL)
instability of the RN black hole without scalar hair~\cite{Myung:2018vug}.
A difference with the ESGB theory~\cite{Silva:2017uqg} is that there is no scalarization bands in the EMS theory,
implying no upper bound on the coupling constant $\alpha$ as the $n=0(\alpha\ge 8.019),1(\alpha\ge 40.84),2(\alpha\ge 99.89),\cdots$ scalarized charged black holes.

The scalarized black holes without charge have been  found from the ESGB
theories~\cite{Silva:2017uqg,Antoniou:2017acq,Doneva:2017bvd}.
It is emphasized  that these black holes with scalar hair  are connected  to the appearance of  instability
for the Schwarzschild black hole without scalar hair.
We note that the instability of Schwarzschild black hole in ESGB theory is considered  as not the tachyonic instability
but the GL instability~\cite{Gregory:1993vy} when comparing it with the GL instability
of the Schwarzschild black hole in the Einstein-Weyl gravity~\cite{Myung:2018iyq}.
Here, the notion of the GL instability comes from the three observations~\cite{Whitt:1985ki,Myung:2013doa,Lu:2017kzi,Stelle:2017bdu}:
i) The instability is based on the $s(l=0)$-mode perturbation for either  massive scalar or massive tensor.
ii) The perturbed equation should include an effective mass term, so that the potential $V(r)$
develops negative region near the horizon of black hole but it becomes positive just after crossing the $r$-axis, leading to $\int^{\infty}_{r_+} dr [V(r)/f(r)] >0$ with the metric function $f(r)$.
Actually, this corresponds to a weaker condition than the sufficient condition of instability ($\int^{\infty}_{r_+} dr [V(r)/f(r)] <0$) including the tachyonic instability
because the integral of potential may be positive.
iii) The instability of a black hole without  hair  is closely related to the appearance of a newly black hole with hair
     where the hair is defined  by  non-zero scalar  outside and on the horizon.

Concerning the stability of scalarized black holes, it turns out that the $n=0$ black hole is stable against all perturbations, while $n=1,2,\cdots$ black holes
are unstable against the $l=0(s$-mode) scalar perturbation in the Einstein-Born-Infeld-scalar theory~\cite{Doneva:2010ke} and
the ESGB theory~\cite{Blazquez-Salcedo:2018jnn}. The former was based on the scalar perturbation only, while the latter was based on the spherically symmetric tensor perturbations including the scalar perturbation. For the stability of scalarized charged black hole in the EMS theory, the $n=0$ black hole was  mentioned within the scalar perturbation~\cite{Herdeiro:2018wub}.

In this work, we wish to carry out the stability analysis on the scalarized  charged black holes in the EMS theory by computing quasinormal mode spectrum.
We wish to employ the full tensor-vector-scalar perturbations splitting into the axial and polar parts.
Observing the potentials around the $n=0,1,2$ black holes with $q=0.7$ and together with computing quasinormal frequencies of the five  physically propagating modes,
we will find that the $n=0$ black hole is stable against all perturbations, while $n=1,2$ black holes
are unstable against the $l=0(s$-mode) scalar perturbation in the EMS theory.

\section{Scalarized charged black holes} \label{sec1}

We start by mentioning  the action of EMS theory without scalar potential~\cite{Herdeiro:2018wub}
\begin{equation}
S_{\rm EMS}=\frac{1}{16 \pi}\int d^4 x\sqrt{-g}\Big[ R-2\partial_\mu \phi \partial^\mu \phi-e^{\alpha \phi^2} F^2\Big],\label{Action1}
\end{equation}
where $\phi$ is a scalar field, $\alpha$ is a Maxwell-scalar coupling constant as a  mass-like parameter,
and $F^2=F_{\mu\nu}F^{\mu\nu}$ is the Maxwell kinetic term. In this work, we do not consider  the  Einstein-Maxwell-dilaton theory with a usual coupling of $e^{\alpha \phi}$~\cite{Garfinkle:1990qj,Brito:2018hjh}.
The EMS theory describes three of a massive scalar, a massless vector, and
a massless tensor which lead to  five (1+2+2=5) physically dynamical modes  propagating on the scalarized charged black hole background.

We  derive  the Einstein  equation from the action (\ref{Action1})
\begin{eqnarray}
 G_{\mu\nu}=2\partial _\mu \phi\partial _\nu \phi -(\partial \phi)^2g_{\mu\nu}+2T_{\mu\nu} \label{equa1}
\end{eqnarray}
with $G_{\mu\nu}=R_{\mu\nu}-(R/2)g_{\mu\nu}$ and  $T_{\mu\nu}=e^{\alpha \phi^2}(F_{\mu\rho}F_{\nu}~^\rho-F^2g_{\mu\nu}/4)$.
The Maxwell equation takes the form
\begin{equation} \label{M-eq}
\nabla^\mu F_{\mu\nu}-2\alpha \phi\nabla^{\mu} (\phi)F_{\mu\nu}=0.
\end{equation}
Importantly,  the scalar equation is given by
\begin{equation}
\square \phi -\frac{\alpha}{2} e^{\alpha \phi^2}F^2 \phi=0 \label{s-equa}.
\end{equation}
For our purpose, we introduce  the metric ansatz as~\cite{Herdeiro:2018wub}
\begin{eqnarray}\label{nansatz}
ds^2=\bar{g}_{\mu\nu}dx^\mu dx^\nu=-N(r)e^{-2\delta(r)}dt^2+\frac{dr^2}{N(r)}+r^2(d\theta^2+\sin^2\theta d\varphi^2)
\end{eqnarray}
with a metric function $N(r)=1-2m(r)/r$, in addition to  $U(1)$ potential  $\bar{A}=v(r)dt$ and scalar $\bar{\phi}(r)$.
We would like to mention that the RN black hole solution [$\tilde{N}(r)=1-2M/r+Q^2/r^2, \delta(r)=\bar{\phi}(r)=0$]  is defined, irrespective of  any value of $\alpha$.
However, a scalarized charged black hole is defined by restricting an allowable range for $\alpha$.
The threshold of  instability for a RN black hole is closely related to the appearance
of the $\alpha\ge 8.019$ fundamental branch which is identified with the $n=0$ scalarized charged  black hole.
Also, the static scalar perturbation around the RN black hole indicates  the appearance of $n=1,2\cdots$ scalarized charged black holes.

First of all, we consider the  static scalar perturbed equation [$(\tilde{\nabla}^2-\alpha \tilde{F}^2/2)\delta\phi=0$] with $\delta \phi=Y_{lm}(\theta,\varphi)\varphi_l(r)$ on the RN black hole background to identify how the $n=0$, 1, 2 black holes come out as
\begin{equation}
\frac{1}{r^2}\frac{d}{dr}\Big[r^2 \tilde{N}(r)\frac{d\varphi_l(r)}{dr}\Big]-\Big[\frac{l(l+1)}{r^2}-\frac{\alpha Q^2}{r^4}\Big] \varphi_l(r)=0
\end{equation}
which describes an eigenvalue problem in the radial direction: for a given $l=0$, requiring an asymptotically vanishing, smooth scalar field
selects a discrete set of $n=0$, 1, 2, $\cdots$. Actually, these determine the bifurcation points of scalar solution
as $\alpha_n(q=0.7)=\{8.019,~40.84,~99.89,\cdots\}$. In Fig. 1, these solutions are classified
by the node number $n$ for $\varphi(z)=\varphi_{l=0}(z)$ with $z=r/(2M)$. Furthermore, $n$ denotes the order number for classifying different branches of scalarized black holes.
\begin{figure*}[t!]
   \centering
   \includegraphics{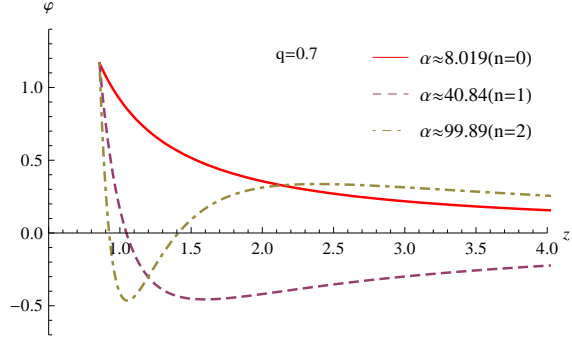}
\caption{Radial profiles of $\varphi(z)=\varphi_{l=0}(z)$ as function of $z=r/(2M)$ for the first three perturbed scalar solutions on the RN black hole with $q=0.7$.
Here $n$ represents the number of nodes for $\varphi(z)$ and it denotes the order number for labeling scalarized black holes on later.  }
\end{figure*}

Now, we  focus on looking for  a scalarized charged black hole with  $q=Q/M=0.7$.
Plugging (\ref{nansatz}) into (\ref{equa1})-(\ref{s-equa}), one has  the four equations
\begin{eqnarray}\label{neom}
&&-2m'(r)+e^{2\delta(r)+\alpha(\bar{\phi}(r))^2}r^2(v'(r))^2+[r^2-2rm(r)](\bar{\phi}'(r))^2=0,\label{neom1}\\
&&\delta'(r)+r(\bar{\phi}'(r))^2=0,\label{neom2}\\
&&v'(r)\Big(2+r\delta'(r)+2r\alpha\bar{\phi}(r)\bar{\phi}'(r)\Big)+r v''(r)=0,\label{neom3}\\
&&e^{2\delta(r)+\alpha(\bar{\phi}(r))^2}r^2\alpha\bar{\phi}(r)(v'(r))^2+r[r-2m(r)]\bar{\phi}''(r)\nonumber\\
&&-\Big(m(r)[2-2r\delta'(r)]
+r[-2+r+2m'(r)]\delta'(r)\Big)\bar{\phi}'(r)=0, \label{neom4}
\end{eqnarray}
where the prime ($'$) denotes differentiation with respect to its argument.
From (\ref{neom3}), one has a relation of  $v'=-e^{-\delta-\alpha \bar{\phi}^2} Q^2/r$.
Considering  an outer horizon located at $r=r_+=0.857$ in the RN black hole,  one finds a numerical solution to four equations in the near-horizon
\begin{eqnarray}\label{nexpr}
&&m(r)=\frac{r_+}{2}+m_1(r-r_+)+\ldots,\label{aps-1}\\
&&\delta(r)=\delta_0+\delta_1(r-r_+)+\ldots,\label{aps-2}\\
&&\bar{\phi}(r)=\phi_0+\phi_1(r-r_+)+\ldots,\label{aps-3}\\
&&v(r)=v_1(r-r_+)+\ldots,\label{aps-4}
\end{eqnarray}
where the coefficients are determined  by
\begin{eqnarray}\label{ncoef}
 m_1=\frac{e^{-\alpha\phi_0^2}Q^2}{2r_+^2},\quad
\delta_1=-r_+\phi_1^2,\quad \phi_1=\frac{\alpha\phi_0 Q^2}{r_+(Q^2-e^{\alpha\phi_0^2}r_+^2)},\quad v_1=-\frac{e^{-\delta_0-\alpha\phi_0^2}Q}{r_+^2}.
\end{eqnarray}
This near-horizon solution involves  two parameters of  $\phi_0=\bar{\phi}(r_+,\alpha)$ and $\delta_0=\delta(r_+,\alpha)$, which will be
determined  by  matching  (\ref{aps-1})-(\ref{aps-4}) with the asymptotic  solution in the far-region
\begin{eqnarray}\label{ncoef}
m(r)&=&M-\frac{Q^2+Q_s^2}{2r}+\ldots,~\bar{\phi}(r)=\frac{Q_s}{r}+\ldots, \nonumber \\
\delta(r)&=&\frac{Q_s^2}{2r^2}+\ldots,~v(r)=\Phi+\frac{Q}{r}+\ldots, \label{insol}
\end{eqnarray}
which include the scalar charge $Q_s$ and the electrostatic potential  $\Phi$.
\begin{figure*}[t!]
   \centering
   \includegraphics{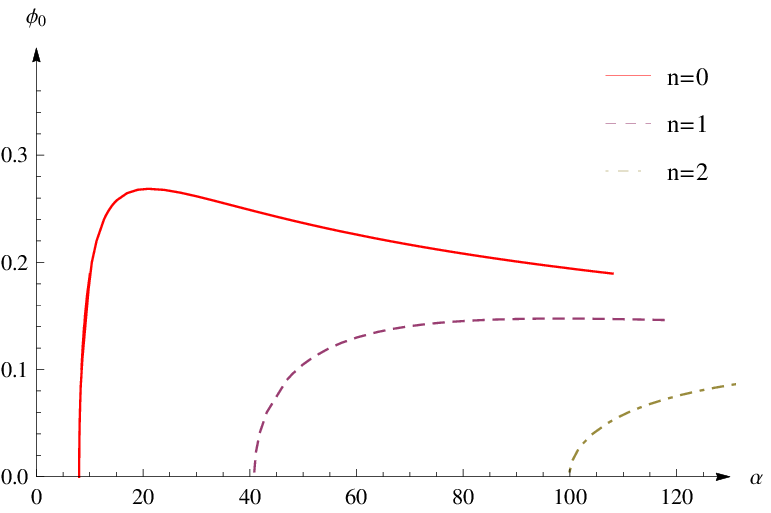}
    \hfill%
    \includegraphics{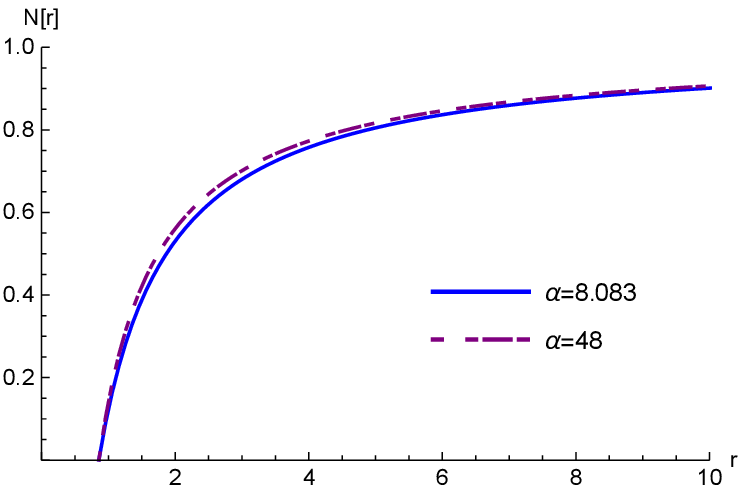}
\caption{(Left) The scalar field $\phi_0=\bar{\phi}(r_+)$ at the horizon as function of $\alpha$. The $n=0$ fundamental branch starts from the first bifurcation point at $\alpha=8.019$,
while $n=1,2$ excited branches  start from the second point at $\alpha=40.84$ and the third point at $n= 99.89$.
(Right) The scalarized charged black hole solutions for the $n=0 (\alpha\ge 8.019)$ fundamental branch. Here we display two metric functions $N(r)$  with $\alpha=8.083$ and 48 residing in the $n=0$  fundamental branch.    }
\end{figure*}

As a concrete scalarized black hole solution with $q=0.7$, we display the two numerical solutions [metric function $N(r)$ only] with the coupling constant $\alpha = 8.083,48$  locating on the $n=0(\alpha \ge 8.019)$ fundamental branch in Fig. 2. It is worth noting that the $n=1(\alpha\ge 40.84)$, $2(\alpha\ge 99.89)$  excited branch solutions  take the similar forms as the $n=0$ case.  For simple notation, we call these scalarized charged black holes as the $n=0,1,2,\cdots$ black holes.

At this stage we mention that  our choice of $q=0.7$ is nothing special, but it is chosen for
a non-extremal black hole between the Schwarzschild ($q=0$) and the extremal  black holes. When the charge $q$ is bigger (smaller) than $q=0.7$, one expects to find similar
solutions and quasinormal modes.
Hence, we will perform the stability analysis on  the $n=0,1,2$ black hole solutions with $q=0.7$ in the next section. Although the $n>2$ black holes exist, it is expected that
they show similar features as the $n=1,2$ black holes show.

\section{Linearized equations}

We consider  the perturbed fields around the background quantities
\begin{eqnarray}
  g_{\mu\nu} =\bar{g}_{\mu\nu}+h_{\mu\nu},~~
  A_\mu =\bar{ A}_\mu+a_{\mu},~~
  \phi = \bar{\phi} +\delta \phi. \label{per-3}
\end{eqnarray}
Plugging (\ref{per-3}) into Eqs.(\ref{equa1})-(\ref{s-equa}) leads to complicated  linearized equations.
Considering  ten degrees of freedom for $h_{\mu\nu}$, four for $a_{\mu}$, and one for $\delta \phi$ initially,
the EMS theory describing a massive scalar and massless vector-tensor propagations provides  five (1+2+2=5) physically propagating modes on the black hole background.
The stability analysis should be based on these physically propagating fields as the solutions to the linearized equations.
In a spherically
symmetric background (\ref{nansatz}), the perturbations can be decomposed into spherical harmonics $Y_{l}^{m}(\theta,\varphi)$ with multipole index
$l$ and azimuthal number $m$. This decomposition
splits  the tensor-vector perturbations into ``axial (A)" which acquires
a factor $(-1)^{l+1}$ under parity inversion and ``polar (P)"
which  acquires a factor $(-1)^l$.

We expand the metric perturbations in tensor spherical
harmonics under the  Regge-Wheeler gauge. For the axial part with two modes $h_0$ and $h_1$,
the perturbed metric takes the form
\begin{eqnarray}
h^{\rm A}_{\mu\nu}(t,r,\theta,\varphi)=\int d\omega e^{-i\omega t}\sum_{l,m}\left[
\begin{matrix}
0&0&-\frac{h_0(r)\partial_\varphi Y_{l}^m}{\sin \theta}&h_0(r)\sin \theta\partial_\theta Y_{l}^m\\
\ast&0&-\frac{h_1(r)\partial_\varphi Y_{l}^m}{\sin \theta}&h_1(r)\sin \theta\partial_\theta Y_{l}^m\\
\ast&\ast&0&0\\
\ast&\ast&\ast&0
\end{matrix}
\right],
\end{eqnarray}
where asterisks denote symmetrization.
For polar perturbations with four modes ($H_0,H_1,H_2,K$), we have
\begin{eqnarray}
h^{\rm P}_{\mu\nu}=\int d\omega e^{-i\omega t}\sum_{l,m}\left[
\begin{matrix}
H_0(r)e^{-2\delta(r)}N(r)&H_1(r)&0&0\\
\ast&\frac{H_2(r)}{N(r)}&0&0\\
\ast&\ast&r^2 K(r)&0\\
\ast&\ast&\ast&r^2\sin^2\theta K(r)
\end{matrix}
\right]Y_{l}^m.
\end{eqnarray}
On the other hand, we decompose  the vector perturbations into
\begin{eqnarray}
a^{\rm A}_{\mu}=\int d\omega e^{-i\omega t}\sum_{l,m}\left[
\begin{matrix}
0,&0,&-\frac{u_4(r)\partial_\varphi Y_{l}^m}{\sin \theta},&u_4(r)\sin \theta\partial_\theta Y_{l}^m
\end{matrix}
\right]
\end{eqnarray}
and
\begin{eqnarray}
a^{\rm P}_{\mu}=\int d\omega e^{-i\omega t}\sum_{l,m}\left[
\begin{matrix}
\frac{u_1(r)Y_{l}^m}{r},&\frac{u_2(r)Y_{l}^m}{rN(r)},&0&0
\end{matrix}
\right],
\end{eqnarray}
where we  gauge  $a^{\rm P}_{\theta,\varphi}$ away.
Lastly, we have a polar scalar perturbation as
\begin{eqnarray}
\delta\phi=\int d\omega e^{-i\omega t}\sum_{l,m}\delta \phi_1(r)Y_{l}^m.
\end{eqnarray}
The linearized equations could be split into axial  and  polar parts.

In general, the axial part is composed of two coupled equations for  Maxwell $\hat{F}(u_4)$ and Regge-Wheeler $\hat{K}(h_0,h_1)$,
\begin{eqnarray} \label{axial-1}
 \Big[\frac{d^2}{dr_{*}^2}+\omega^2\Big]\hat{F}(r)&=&V^{\rm A}_{\rm FF}(r)\hat{F}(r)+V^{\rm A}_{\rm FK}(r)\hat{K}(r), \\
 \Big[\frac{d^2}{dr_{*}^2}+\omega^2\Big]\hat{K}(r)&=&V^{\rm A}_{\rm KF}(r)\hat{F}(r)+V^{\rm A}_{\rm KK}(r)\hat{K}(r),\label{axial-2}
\end{eqnarray}
where the potentials are given by
\begin{eqnarray}
  V^{\rm A}_{\rm FF}(r) &=& \frac{N}{r^2 e^{2\delta}}\Big[e^{2\delta+\alpha\bar{\phi}^2}r^2(4-\alpha\bar{\phi}^2)(v')^2+l(l+1)
  +\alpha rN\bar{\phi}'\left(r(1+\alpha\bar{\phi}^2)\bar{\phi}'-2\bar{\phi}\right)\Big], \\
  V^{\rm A}_{\rm FK}(r) &=& V^{\rm A}_{\rm KF}(r)=-\frac{2e^{-\delta+\alpha\bar{\phi}^2/2}(l-1)(l+2)N v'}{r}, \\
  V^{\rm A}_{\rm KK}(r) &=& \frac{N}{r^2 e^{2\delta}}\Big[(l-1)(l+2)-r N'+N(2+r\delta)\Big].
\end{eqnarray}
Here the tortoise coordinate $r_*\in(-\infty,\infty)$ is defined by the relation of $dr_*/dr= e^\delta/N$.
At this stage, it is worth noting that in the limits of $\bar{\phi}=\delta=0$, $V^{\rm A}_{\rm FF}(r)$, $V^{\rm A}_{\rm FK}(r)$, and $ V^{\rm A}_{\rm KK}(r)$ recovers those for the RN black hole in the EM theory~\cite{Chandrasekhar:1979iz}. In addition, we would like to mention that  the diagonalized forms may  be adopted to compute quasinormal modes propagating around scalarized charged black holes.
 However, it is not easy to find a simple method to diagonalize  two coupled equations (\ref{axial-1}) and (\ref{axial-2}). Actually, the diagonalization is not easily performed because of the presence of the background scalar.  Instead, we will derive the quasinormal modes propagating around scalarized charged black holes by solving the two coupled equations directly.

On the other hand, the polar part is composed of  six coupled equations for Zerilli, Maxwell,  and scalar  as
\begin{eqnarray}
 K'(r)&=&-\left(\frac{l(l+1)+2N+2r N'-2}{2r^2}+e^{2\delta+\alpha\bar{\phi}^2}v'^2+N\phi'(r)^2\right)H_1(r)\label{pol-eq1}\\
 &&\frac{H_0(r)}{r}+\left(\frac{N'}{2N}-\frac{1}{r}-\delta'\right)K(r)-\frac{2\bar{\phi}'}{r}\delta \phi_1(r), \nonumber\\
 H_1'(r)&=&-\frac{4ie^{\alpha\bar{\phi}^2}v'}{\omega}f_{12}(r)-\frac{H_0(r)+K(r)}{N}+\left(\delta'-\frac{N'}{N}\right)H_1(r), \label{pol-eq2}\\
H_0'(r)&=&\left(\frac{1}{r}+2\delta'-\frac{N'}{N}\right)\Big[H_0(r)-K(r)\Big]
+\frac{4e^{2\delta+\alpha\bar{\phi}^2}v'}{N}f_{02}(r)+\frac{2\bar{\phi}'}{r}\delta\phi_1(r) \label{pol-eq3}\\
&&+\left(\frac{e^{2\delta}\omega^2}{N}-e^{2\delta+\alpha\bar{\phi}^2}v'^2-N\phi'^2-\frac{l(l+1)}{2r^2}-\frac{N+rN'-1}{r^2}\right)H_1(r),\nonumber\\
f_{02}'(r)&=&v'K(r)+\frac{2\alpha\bar{\phi}V'}{r}\delta\phi_1(r)+\left(\frac{l(l+1)ie^{-2\delta}N}{r^2\omega}-i\omega\right)f_{12}(r),\label{pol-eq4}\\
f_{12}'(r)&=&-\frac{i\omega e^{2\delta}}{N^2}f_{02}(r)+\left(\delta'-2\alpha\bar{\phi}\bar{\phi}'-\frac{N'}{N}\right)f_{12}(r),\label{pol-eq5}\\
  \delta\phi''_1(r)&=&\Big[\frac{l(l+1)}{r^2N}-\frac{e^{2\delta}\omega^2}{N^2}
+\frac{N'+e^{2\delta+\alpha\bar{\phi}^2}r\alpha(2\alpha\bar{\phi}^2-1)v'^2-N(\delta'-4r\bar{\phi}'^2)}{rN}\Big]\delta\phi_1(r)\nonumber\\
&&+\left(\delta'-\frac{N'}{N}\right)\delta\phi'_1(r)+\frac{2ie^{\alpha\bar{\phi}^2}\alpha\bar{\phi}v'}{r\omega}f_{12}(r)+\frac{4e^{2\delta+\alpha\bar{\phi}^2}r v'\bar{\phi}}{N}f_{02}(r)\label{pol-eq6}\\
&&-\frac{r\left(e^{2\delta+\alpha\bar{\phi}^2}\alpha\bar{\phi}v'^2+(N'-2N\delta')\bar{\phi}'\right)}{N}H_0(r)
+\frac{2re^{2\delta+\alpha\bar{\phi}^2}\alpha\bar{\phi}v'^2}{N}K(r).\nonumber
\end{eqnarray}
Here we have  $H_2(r)=H_0(r)$,
  $f_{12}(r)=\frac{u_2(r)}{rN(r)}$ and $f_{02}(r) =\frac{u_1(r)}{r}$.
Interestingly, these coupled equations describe three physically propagating modes.

\section{Stability Analysis}
The stability analysis will be performed by getting quasinormal frequency of $\omega=\omega_{r}+i\omega_i$ when  solving the linearized equations with appropriate boundary conditions at the outer horizon: ingoing waves and at infinity: purely outgoing waves. Also, the late-time signals from  perturbed black holes are dominated by the fundamental quasinormal mode, which corresponds to the mode with smallest imaginary component. We will compute the lowest quasinormal modes of the scalarized black holes by making use of the
fully numerical background and the linearized equations (\ref{axial-1})-(\ref{axial-2}) for axial part and the linearized equations (\ref{pol-eq1})-(\ref{pol-eq6}) for polar part.
To compute the quasinormal modes, we use a direct-integration method~\cite{Blazquez-Salcedo:2016enn}.

Usually, a positive definite potential $V(r)$ without any negative region guarantees the stability of black hole.
On the other hand, a sufficient condition for instability is given by $\int^{\infty}_{r_+} dr [e^\delta V(r)/N(r)] <0$~\cite{Dotti:2004sh} in accordance with the existence of the unstable modes.
However, some potentials with negative region  near the outer horizon whose integral is positive ($\int^{\infty}_{r_+} dr [e^\delta V(r)/N(r)] >0$) do not imply a definite instability. To determine the instability of the $n=0,1,2$ black holes clearly, one has to solve all linearized equations for physical perturbations numerically.

Accordingly, the criterion to determine whether a black hole is stable or not against the physical perturbations is whether the time evolution  $e^{-i\omega t}$  of the perturbation is decaying or not. If $\omega_i<0(>0)$, the black hole is stable (unstable), irrespective of any value of $\omega_r$.
However, it is a nontrivial task to carry out the stability of a scalarized charged black hole because this black hole comes out as  not an analytic solution but a  numerical solution.
In order to develop the stability analysis, it is convenient to  classify the linearized equations according to multipole index $l=0,1,2,\cdots$ because  $l$ determines number of physical fields at the axial and polar sectors.

\subsection{$l=0$ case: one DOF}
\begin{figure*}[t!]
   \centering
   \includegraphics{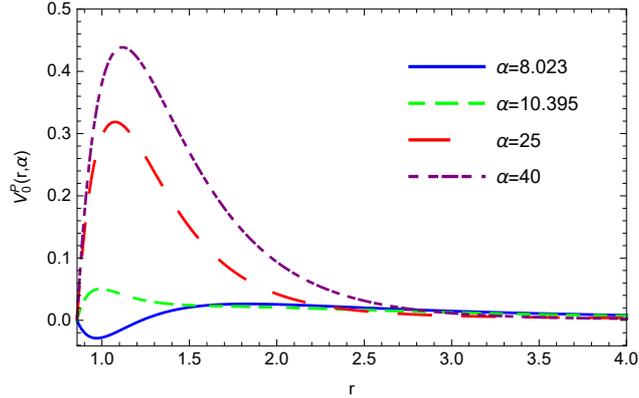}
\caption{Four scalar potential graphs $V^{\rm P}_0(r,\alpha)$  with $l=0$ around the $n=0(\alpha\ge 8.019)$ black hole. The whole potentials are positive definite except that the $\alpha=8.083$ case having negative region near the horizon.}
\end{figure*}

\begin{figure*}[t!]
   \centering
   \includegraphics{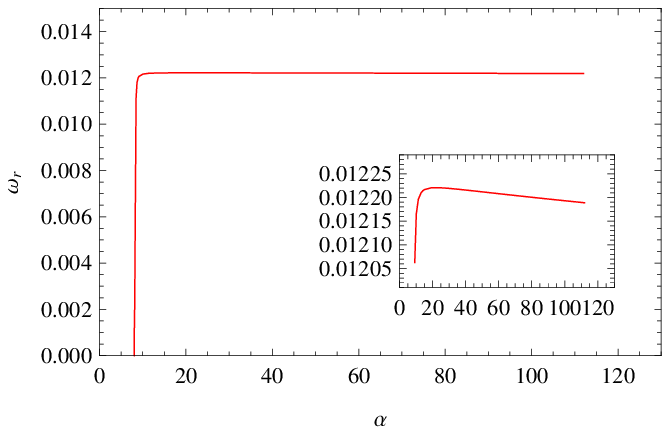}
   \hfill%
   \includegraphics{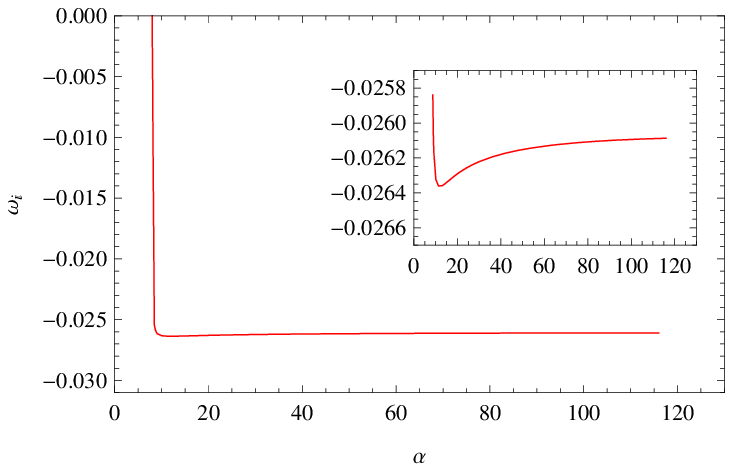}
\caption{ (Left) Real frequency $\omega_r$  and (Right) imaginary frequency $\omega_i$ for a scalar quasinormal mode  with $l=0$ as a function of
$\alpha$ around the $n=0$ black hole.  These start from $\alpha=8.019$. The magnifications of the enclosed regions indicate  the tendency for decreasing and increasing with respect to $\alpha$. }
\end{figure*}

For $l=0$($s$-mode), the linearized equation obtained from the polar part is given entirely  by a scalar equation ($\hat{S}^{\rm P}_0=r \delta \phi_1$)
\begin{equation}
\Big[\frac{d^2}{dr_{*}^2}+\omega^2\Big]\hat{S}^{\rm P}_0-V^{\rm P}_{0}(r,\alpha)\hat{S}^{\rm P}_0=0,
\end{equation}
where the potential $V^{\rm P}_{\rm 0}(r,\alpha)$ is given by~\cite{Herdeiro:2018wub}
\begin{equation}
 V^{\rm P}_{\rm 0}(r,\alpha)=\frac{N}{e^{2\delta}r^2}\Big[1-N-2r^2(\bar{\phi}')^2+e^{-\alpha \bar{\phi}^2}Q^2\Big(\frac{2(-\alpha\bar{\phi}+r\bar{\phi}')^2-\alpha-1}{r^2}\Big)\Big].
 \label{potl=0}
\end{equation}
We display four scalar potentials $V^{\rm P}_{\rm 0}(r,\alpha)$  in Fig. 3 for $l=0$ case around the $n=0$ black hole. The whole potentials are positive definite except that the $\alpha=8.083$ case having negative region near the horizon does not  represent instability really because it is near the threshold of instability. Actually,  the $n=0$ black hole is stable against the $l=0$($s$-mode) scalar perturbation since the $n=0$ case corresponds to the threshold of instability satisfying the condition of $\int^{\infty}_{r_+} dr [e^\delta V(r)/N(r)] >0$. Although this condition does not rule out the possibility of unstable modes, one does not find any unstable modes.
 We confirm it  from Fig. 4 that  the imaginary frequency $\omega_i$ is negative  for $\alpha \ge 8.019$, implying a stable $n=0$ black hole.  We observe that although $\omega_r$ and $\omega_i$ seem to be  independent of $\alpha$, it is not true. The magnifications of the enclosed regions show the tendency for decreasing and increasing with respect to $\alpha$.

\begin{figure*}[t!]
   \centering
   \includegraphics{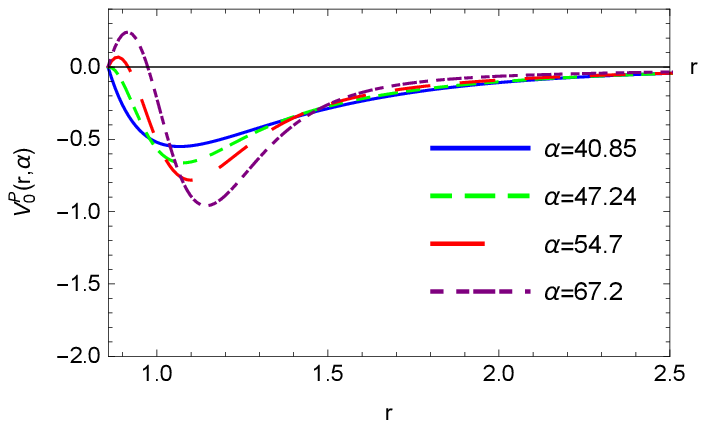}
   \hfill%
   \includegraphics{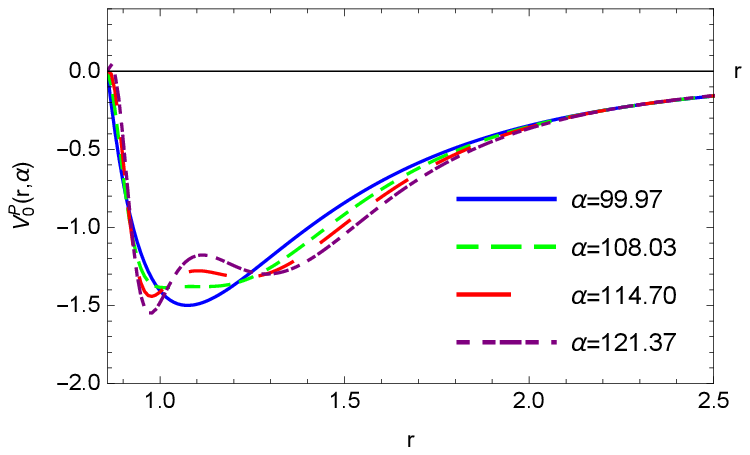}
\caption{Four scalar potential graphs $V^{\rm P}_0(r,\alpha)$  with $l=0$ around  (Left) $n=1(\alpha\ge 40.84)$ black hole and  (Right) $n=2 (\alpha\ge 99.89)$ black hole. }
\end{figure*}
\begin{figure*}[t!]
   \centering
   \includegraphics{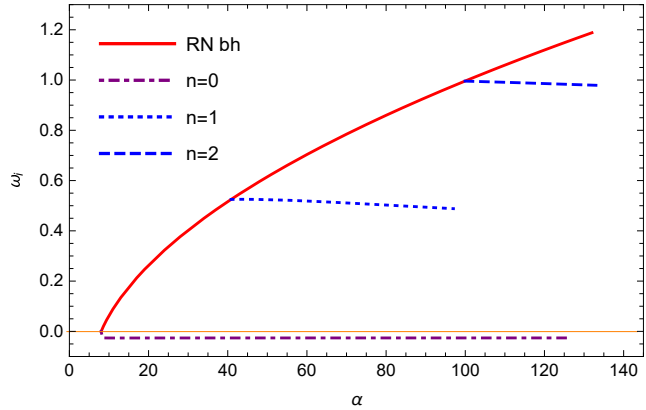}
\caption{The positive imaginary frequency $\omega_i$ ($\omega_r=0$) as function of $\alpha$ for  the $l=0$ scalar mode around
the $n=1,2$ black holes. A red solid curve with $q=0.7$ represents the quasinormal frequency of $l=0$ scalar  as function of $\alpha$ around the RN black hole~\cite{Myung:2018vug}, showing the instability of RN black holes.
The red solid curve starts from the first bifurcation point at $\alpha=8.019$. Attaching (Right) Fig.4 on Fig. 6 shows the negative imaginary frequency around the $n=0$ black hole clearly. }
\end{figure*}
Now let us turn to the stability issue of the $n=1,2$  black holes.
We observe from Fig. 5 that  $\int^{\infty}_{r_+} dr [e^\delta V(r)/N(r)] <0$ for the $n=1$ black hole, while all potentials are negative definite for the $n=2$ black hole.
This suggests that the $n=1,2$ black holes are unstable against the $l=0(s$-mode) scalar perturbation.
Clearly, the instability could be found  from Fig. 6 because their imaginary frequencies are positive.
Here, the red curve denotes the instability (positive $\omega_i$) of RN black hole as a function of $\alpha$.
Attaching (Right) Fig.4 on Fig. 6 indicates the negative imaginary frequency around the $n=0$ (stable) black hole.  This instability may be  regarded as the GL instability because it corresponds to the $s$-mode instability. Actually, Fig. 6 is regarded as  our main result  to show the (in)stability of $n=0$, 1, 2 black holes.

Hereafter, we will perform  the stability analysis for higher multipoles on the $n=0$ black hole only because the $n=1,$ 2 black holes turned out to be unstable against the $l=0(s)$-mode perturbation. In other words, it is not meaningful to carry out a further analysis for the unstable $n=1,$ 2 black holes.

\subsection{$l=1$ case: three DOF}

\begin{figure*}[t!]
   \centering
   \includegraphics{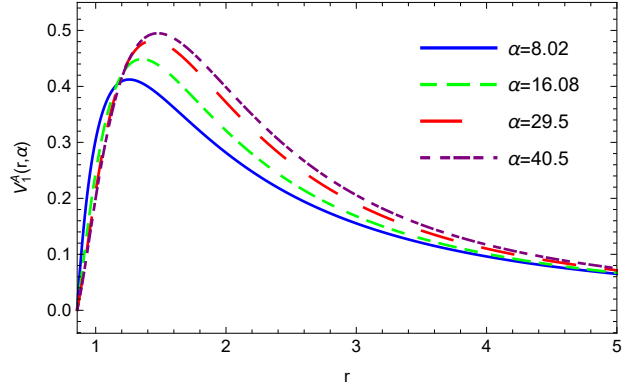}
\caption{The positive  potential $V^{\rm A}_1(r,\alpha\ge 8.019)$  for axial  $l=1$ vector perturbation propagating around the  $n=0$ black hole.}
\end{figure*}

\begin{figure*}[t!]
   \centering
   \includegraphics{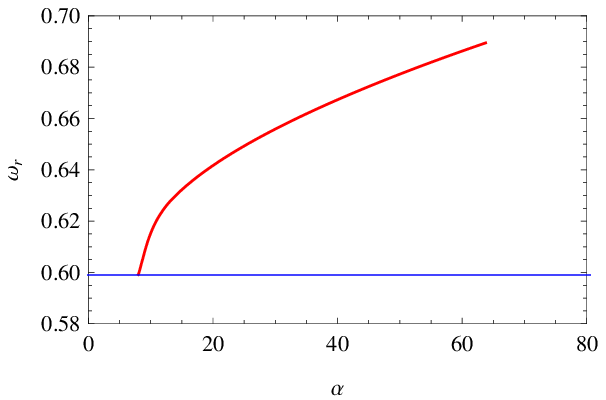}
    \hfill%
    \includegraphics{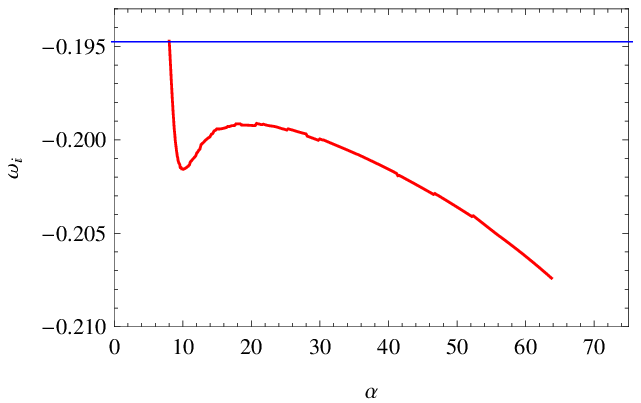}
\caption{(Left) Real frequency  and (Right) negative imaginary frequency  as function of $\alpha$ for  axial  $l=1$ vector mode  around the  $n=0$ black hole.
 At $\alpha=8.019$, one recovers the fundamental quasinormal frequency for the $l=1$ EM  mode around RN black hole (blue horizontal lines).
}
\end{figure*}

For $l=1$ case, the axial linearized equation is given by
\begin{equation}
\Big[\frac{d^2}{dr_{*}^2}+\omega^2\Big]\hat{F}-V^{\rm A}_1(r,\alpha)\hat{F}=0,
\end{equation}
where the potential takes the form
\begin{eqnarray}
V_1^{\rm A}(r,\alpha)=-\frac{e^{-2\delta} N}{r^2}\Big[&N&\Big(4-\alpha^2\bar{\phi}^2+\alpha r(\bar{\phi}^2)'-r^2(\alpha-4+2\alpha^2\phi^2)(\bar{\phi}')^2\Big)\nonumber \\
&-&6+4r N'+\alpha^2\bar{\phi}^2(1-r N')\Big]\label{potl=1a}
\end{eqnarray}
We note that in the limits of $\bar{\phi}(r)\to 0$ and $\delta \to 0$, Eq.(\ref{potl=1a}) reduces to the axial vector perturbed
equation in the Einstein-Maxwell (EM) theory~\cite{Leaver:1990zz,Berti:2005eb}
\begin{equation}
V^{\rm A}_{\rm 1EM}(r)=-\frac{N}{r^2}\Big(4N-6+4r N'\Big).\label{potl=1a1}
\end{equation}
We find from Fig. 7 that all potentials are  positive definite for the $n=0$ black hole.
 This means that the $n=0$ black hole is stable against the axial $l=1$ vector perturbation.
 We confirm it  from Fig. 8 that  $\omega_i$ is  negative, indicating a stable black hole.
 Moreover, it is interesting to note that the quasinormal frequency at $\alpha=8.019$  coincides with  that for the $l=1$ fundamental EM mode ($0.59896-0.19475i$) around the RN black hole~\cite{Cho:2011sf,Matyjasek:2017psv}.

Finally, we  find  the vector-led and scalar-led modes around the $n$=0 black hole from  the polar $l=1$  linearized equations (\ref{pol-eq1})-(\ref{pol-eq6}).
We find from Fig. 9 that all $\omega_i$ of these modes around the $n=$0 are negative, implying a stable black hole.
\begin{figure*}[t!]
   \centering
   \includegraphics{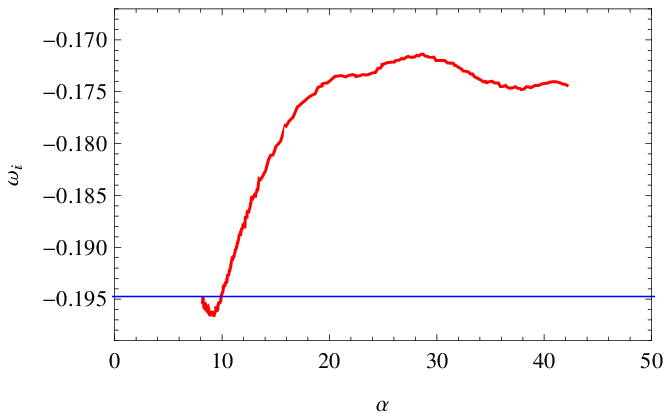}
    \hfill%
    \includegraphics{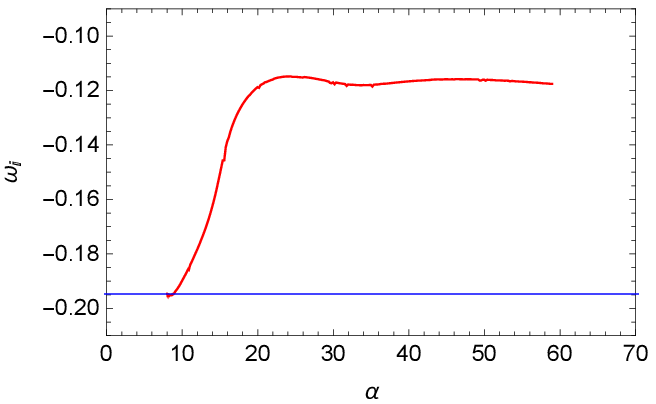}
\caption{Imaginary frequencies  as function of $\alpha$ for polar  $l=1$ vector-led mode (Left) and scalar-led mode (Right) around the  $n=0$  black hole.
}
\end{figure*}

\subsection{$l=2$ case: five DOF}
First of all, we consider the axial part because of its simplicity.
The axial  linearized equations are given by   two coupled equations for Regge-Wheeler-Maxwell system (\ref{axial-1}) and (\ref{axial-2}) with $l=2$.
Solving these coupled equation with boundary conditions leads to negative quasinormal frequencies $\omega_i<0$ for $l=2$ vector-led and gravitational-led modes  around  the $n=0$ black hole (see Fig. 10), implying stable black hole. Here we find the  fundamental frequency of $1.07302-0.197542i$ for the $l=2$ vector-led mode around the RN black hole in the EM theory~\cite{Cho:2011sf,Matyjasek:2017psv}.
We note that the $l=2$ fundamental frequency of  $0.784997-0.179809i$ (for  gravitational-led mode around the RN black hole in the EM theory)
plays the role of  a starting point for the $n=0$ black hole.
\begin{figure*}[t!]
   \centering
   \includegraphics{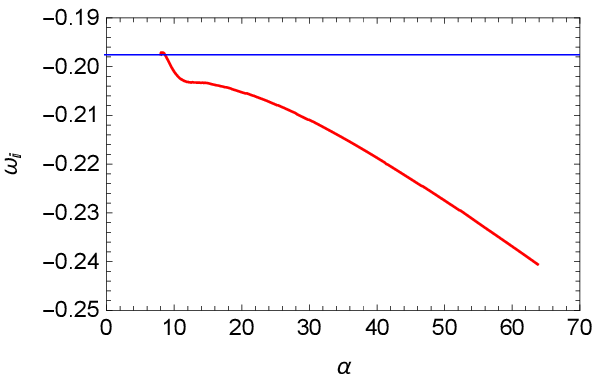}
    \hfill%
    \includegraphics{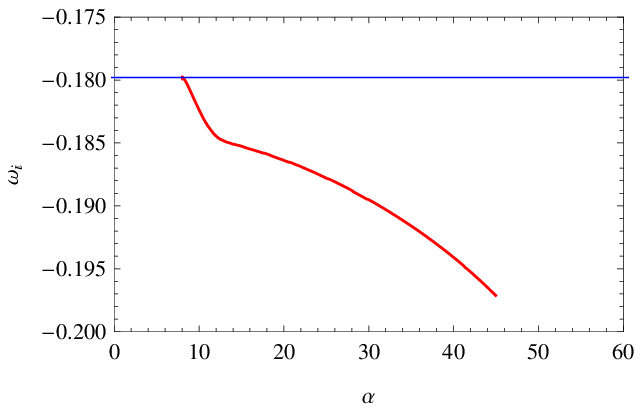}
\caption{Imaginary frequencies  as function of $\alpha$ for  axial  $l=2$ vector-led mode (Left) and  gravitational-led mode (Right) around the  $n=0$  black hole.
}
\end{figure*}

Now, the polar  $l=2$ linearized equations are given by Eqs.(\ref{pol-eq1})-(\ref{pol-eq6}) with $l=2$.
Here we have three modes: vector-led, gravitational-led, and scalar-led modes.
We find from Figs. 11 and 12  that all $\omega_i$ are negative, implying the stable $n=0$ black hole.
It is worth noting that the $l=2$ fundamental frequencies of vector-led and gravitational-led modes  around the RN black hole in the EM theory  take the same values as in the axial case~\cite{Cardoso:2016olt}.
For the polar $l=2$  scalar-led mode,  the quasinormal  frequency  starts  from $0.9923-0.1834i$ for $\alpha=8.019$.

\begin{figure*}[t!]
   \centering
   \includegraphics{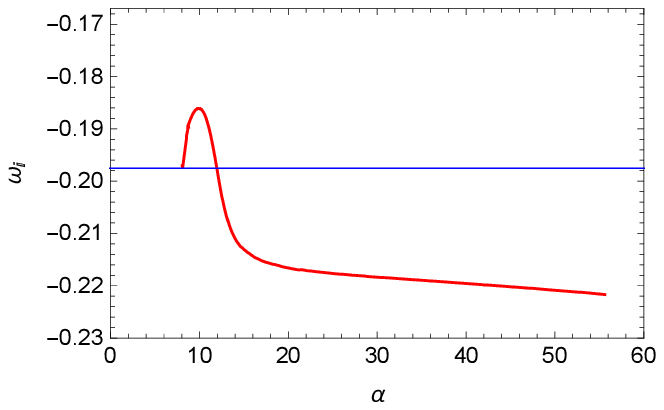}
    \hfill%
    \includegraphics{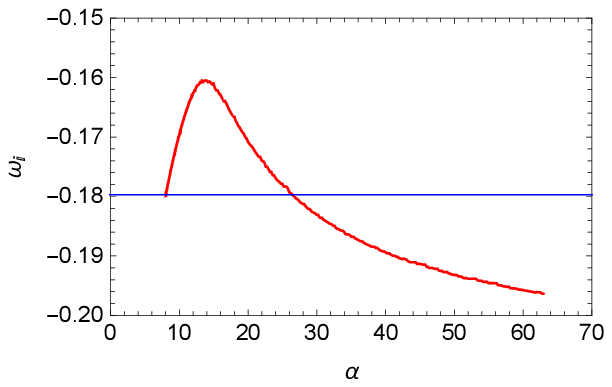}
\caption{Imaginary frequencies  as function of $\alpha$ for polar  $l=2$ vector-led mode (Left) and gravitational-led mode (Right) around the  $n=0$ black hole. }
\end{figure*}
\begin{figure*}[t!]
   \centering
     \includegraphics{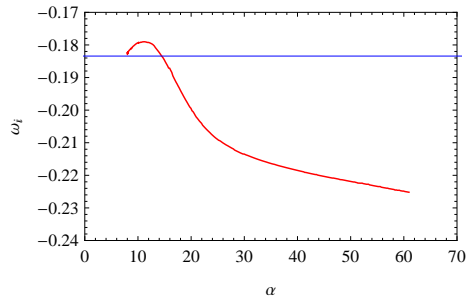}
\caption{Imaginary frequency  for  polar  $l=2$ scalar-led mode  around the  $n=0$ black hole.}
\end{figure*}
\section{Summary}

In this work, we performed the stability analysis of the scalarized charged black holes in the EMS theory by computing quasinormal mode spectrum.
This  is a nontrivial task and completing it  takes a long time because these  black holes are found in numerically.

We have shown that  the $n=1(\alpha\ge 40.84),2(\alpha\ge 99.89)$ excited black holes are unstable against  the $s(l=0)$-mode scalar perturbation only, while the $n=0(\alpha \ge 8.019)$ fundamental black hole is stable against all scalar-vector-tensor perturbations.
In the former case, the instability of the $n=1,2,\cdots$ black holes is regarded as the Gregory-Laflamme instability  because it arose from the $s(l=0)$ mode with an effective mass term.
In the latter, we found negative quasinormal frequencies ($\omega_i<0$) of $9=1(l=0)+3(l=1)+5(l=2)$ physical modes around $n=0$ black hole.
We could not find any unstable modes from the $l=0,1,2$ scalar-vector-tensor  perturbations around the $n=0$ black hole, as in the RN black hole~\cite{Myung:2018vug}.
Even though we have carried out the stability analysis on the $n=0$, 1, 2 black holes, we expect to find  from Fig. 5 that the other  higher excited ($n=$3, 4, 5,$\cdots$) black holes are unstable against the $s(l=0)$-mode scalar perturbation.
This picture is consistent with other scalarized black holes without charge found in the ESGB theory by making use of spherically symmetric perturbations~\cite{Blazquez-Salcedo:2018jnn}.

 \vspace{1cm}

{\bf Acknowledgments}

We are  grateful to Yunqi Liu for useful discussions. This work was supported by the National Research Foundation of Korea (NRF) grant funded by the Korea government (MOE)
 (No. NRF-2017R1A2B4002057).
 
\end{document}